\documentclass[10pt]{scrartcl}

\PassOptionsToPackage{ps2pdf}{hyperref}
\usepackage{jheppub}

\usepackage{amsfonts}

\def\unity{{\bf 1}}
\def\zero{{\bf 0}}
\newcommand{\I}{{i\mkern1mu}}
\newcommand{\E}{\mathrm{e}}
\newcommand{\Real}{\Re\hspace{-1pt}\mathfrak{e}}
\newcommand{\Realtilde}{\tilde{\Re\hspace{2pt}}\hspace{-3pt}\mathfrak{e}}
\newcommand{\MZ}{\mathbf{M}^{\mathbf{Z}}}
\newcommand{\TZ}{T^Z}
\newcommand{\st}{\tilde{t}}
\newcommand{\sbottom}{\tilde{b}}

\begin{document}


\thispagestyle{empty}

\def\thefootnote{\fnsymbol{footnote}}

\begin{flushright}
MPP--2014--11
\end{flushright}

\vspace{2cm}

\begin{center}

{\large\sc {\bf Two-loop top-Yukawa-coupling corrections to the}}

\vspace{0.4cm}

{\large\sc {\bf   Higgs boson masses in the complex MSSM}}

\vspace{1cm}

Wolfgang Hollik and  
Sebastian Pa{\ss}ehr

\vspace*{.7cm}

{\sl
Max-Planck-Institut f\"ur Physik \\
(Werner-Heisenberg-Institut) \\
F\"ohringer Ring 6, 
D--80805 M\"unchen, Germany
}

\end{center}

\vspace*{2cm}

\begin{abstract}
{}
Results for the leading two-loop corrections of
$\mathcal{O}{\left(\alpha_t^2\right)}$ 
from the Yukawa sector to the Higgs-boson masses
of the MSSM with complex parameters are presented. 
The corresponding self-energies and their renormalization
have been obtained in a Feynman-diagrammatic approach.
A numerical analysis of the new contributions is performed
for the mass of the lightest Higgs boson, supplemented by
the full one-loop result and the  
$\mathcal{O}{\left(\alpha_{t}\alpha_s\right)}$ terms
including complex phases.
In the limit of the real MSSM a previous result is confirmed.
\end{abstract}

\def\thefootnote{\arabic{footnote}}
\setcounter{page}{0}
\setcounter{footnote}{0}

\newpage

\section{Introduction}
The discovery of a new boson~\cite{Aad:2012tfa,Chatrchyan:2012ufa} 
with a mass around 125.6 GeV by the experiments ATLAS and CMS
at CERN has triggered an intensive investigation to reveal the nature
of this particle as a Higgs boson from the mechanism  of electroweak symmetry breaking.
Within the present experimental uncertainties, which are still considerably large,
the measured properties of the new boson 
are consistent with the corresponding predictions for
the Standard Model Higgs boson~\cite{Landsberg:2013ina},
but still a large variety of other interpretations is possible
which are connected to physics beyond the Standard Model. 
Within the theoretical well motivated 
minimal supersymmetric Standard Model~(MSSM),
the observed particle could be classified as a light state within
a richer predicted spectrum. The Higgs sector of the MSSM
consists of two complex scalar doublets leading
to five physical Higgs bosons and three (would-be) Goldstone bosons.
At the tree-level, the physical states are given by  
the neutral $CP$-even $h,H$ and  $CP$-odd $A$ bosons, 
together with the charged $H^{\pm}$ bosons, and can be
parametrized in terms of the $A$-boson mass $m_A$ and
the ratio of the two vacuum expectation values, 
$\tan\beta = v_2/v_1$.
In the MSSM with complex parameters, the cMSSM,
$CP$ violation is induced in the Higgs sector
by loop contributions with complex parameters from other SUSY sectors
leading to mixing between $h,H$ and $A$ in the 
mass eigenstates~\cite{Pilaftsis:1998pe}.
 
Masses and mixings in the neutral sector are sizeably influenced by
loop contributions,  and accordingly intensive work has been invested 
into higher-order calculations of the mass spectrum from the SUSY parameters,
in the case of the real
MSSM~\cite{Heinemeyer:1998jw,Heinemeyer:1998np,Heinemeyer:1999be,Heinemeyer:2004xw,Zhang:1998bm,Espinosa:2000df,Brignole:2001jy,Casas:1994us,Degrassi:2002fi,Heinemeyer:2004gx,Allanach:2004rh,Martin:2001vx}
as well as the cMSSM~\cite{Demir:1999hj,Pilaftsis:1999qt,Carena:2000yi,Heinemeyer:2007aq}. 
The largest loop contributions arise from the Yukawa sector with the
large top Yukawa coupling~$h_t$, or $\alpha_t=h_t^2/(4\pi)$, respectively.
The class of leading two-loop Yukawa-type corrections  of $\mathcal{O}{\left(\alpha_{t}^{2}\right)}$ 
has been calculated so far only in the case of real parameters ~\cite{Espinosa:2000df,Brignole:2001jy}, 
applying the effective-potential method.
Together with the full one-loop result~\cite{Frank:2006yh} and the leading 
$\mathcal{O}{\left(\alpha_{t}\alpha_{s}\right)}$ terms~\cite{Heinemeyer:2007aq}, 
both accomplished in the Feynman-diagrammatic approach including complex parameters, 
it has been implemented in the public program 
{\tt FeynHiggs}~\cite{Heinemeyer:1998np,Degrassi:2002fi,Frank:2006yh,Heinemeyer:1998yj,Hahn:2009zz}.
A calculation of the $\mathcal{O}{\left(\alpha_{t}^{2}\right)}$ terms
for the cMSSM, however, has been missing until now.

In this letter  we present this class of 
$\mathcal{O}{\left(\alpha_{t}^{2}\right)}$ contributions extended to the case
of complex parameters.   The computation has been
carried out in the Feynman-diagrammatic approach; for the special case
of real parameters we obtain a result equivalent to the one in~\cite{Brignole:2001jy}
in an independent way, serving thus as a cross check and as a consolidation
of former spectrum calculations and associated tools.
These new contributions will be included in the code {\tt FeynHiggs}.

\section{Higgs boson masses in the cMSSM}
\subsection{Tree-level relations}
The two scalar $SU(2)$ doublets can be decomposed according to 
\begin{align}
  \label{eq:Higgsfields}
  \mathcal{H}_{1} &= \begin{pmatrix} H_{11}\\ H_{12} \end{pmatrix} = \begin{pmatrix} v_{1} + \frac{1}{\sqrt{2}}(\phi_{1} - \I \chi_{1})\\ -\phi^{-}_{1}\end{pmatrix},&
  \mathcal{H}_{2} &= \begin{pmatrix} H_{21}\\ H_{22} \end{pmatrix} = \E^{\I \xi}\begin{pmatrix} \phi^{+}_{2}\\ v_{2} + \frac{1}{\sqrt{2}}(\phi_{2} + \I \chi_{2})\end{pmatrix} ,
\end{align}
leading to the Higgs potential written as an expansion  
in terms of the components
$\Big[$\,with the notation $\phi^{-}_{1} = \left(\phi^{+}_{1}\right)^{\dagger}, \, \phi^{-}_{2} = \left(\phi^{+}_{2}\right)^{\dagger}\Big]$,
\begin{align}
  \begin{split}
    V_{H} &= -T_{\phi_{1}} \phi_{1} - T_{\phi_{2}} \phi_{2} - T_{\chi_{1}} \chi_{1} - T_{\chi_{2}} \chi_{2}\\
         &\quad + \frac{1}{2}\begin{pmatrix} \phi_{1}, & \phi_{2}, & \chi_{1}, & \chi_{2} \end{pmatrix}
            \begin{pmatrix}\mathbf{M}_{\phi} & \mathbf{M}_{\phi\chi}\\ \mathbf{M}_{\phi\chi}^{\dagger} & \mathbf{M}_{\chi} \end{pmatrix}
            \begin{pmatrix} \phi_{1}\\ \phi_{2}\\ \chi_{1}\\ \chi_{2}\end{pmatrix}
            + \begin{pmatrix} \phi^{-}_{1}, & \phi^{-}_{2}\end{pmatrix} \mathbf{M}_{\phi^{\pm}} \begin{pmatrix} \phi^{+}_{1}\\ \phi^{+}_{2}\end{pmatrix} + \dots  \, ,
  \end{split}
\end{align}
where higher powers in field components have been dropped.
The explicit form of the tadpole coefficients $T_i$ and of the mass matrices $\mathbf{M}$ 
can be found in Ref.~\cite{Frank:2006yh}. They are parametrized by the
phase $\xi$,  the real SUSY breaking quantities 
$m_{1,2}^{2} = \tilde{m}_{1,2}^{2} + \lvert\mu\rvert^{2}$,
and the complex SUSY breaking quantity $m_{12}^{2}$. 
The latter can be redefined as real~\cite{Dimopoulos:1995kn} 
with the help of a Peccei--Quinn transformation~\cite{Peccei:1977hh} 
leaving only the phase $\xi$ as a source of $CP$ violation at the tree-level. 
The requirement of minimizing $V_{H}$ at the vacuum
expectation values $v_{1}$ and $v_{2}$  induces vanishing tadpoles at tree level, 
which in turn leads to $\xi = 0$. As a consequence, also $\mathbf{M}_{\phi\chi}$ 
is equal to zero and $\phi_{1,2}$ are decoupled from $\chi_{1,2}$ at the tree-level. 
The remaining 2$\times$2 matrices  $\mathbf{M}_{\phi}, \mathbf{M}_{\chi}$, $\mathbf{M}_{\phi^{\pm}}$ 
can be transformed into the mass eigenstate basis with the help of
unitary matrices~$D(x) = \big(\begin{smallmatrix} -s_{x} & c_{x}\\ c_{x} & s_{x}\end{smallmatrix}\big)$, 
writing $s_{x} \equiv \sin{x}$ and $c_{x} \equiv \cos{x}$:
\begin{align}
  \label{eq:higgsmixing}
  \begin{pmatrix} h\\ H \end{pmatrix} &= D(\alpha) \begin{pmatrix} \phi_{1}\\ \phi_{2}\end{pmatrix},&
  \begin{pmatrix} A\\ G \end{pmatrix} &= D(\beta) \begin{pmatrix} \chi_{1}\\ \chi_{2}\end{pmatrix},&
  \begin{pmatrix} H^{\pm}\\ G^{\pm} \end{pmatrix} &= D(\beta) \begin{pmatrix} \phi^{\pm}_{1}\\ \phi^{\pm}_{2}\end{pmatrix}.
\end{align}
The Higgs potential in this basis can be expressed as follows,
\begin{align}
\label{eq:HiggsPotential}
  \begin{split}
    V_{H} &= -T_h \, h- T_H \, H - T_A \, A - T_G\,  G\\
         &\quad + \frac{1}{2}\begin{pmatrix} h, & H, & A, & G \end{pmatrix}
            \mathbf{M}_{hHAG}
            \begin{pmatrix} h \\ H\\ A\\ G\end{pmatrix}
            + \begin{pmatrix} H^{-}, & G^{-}\end{pmatrix} 
                \mathbf{M}_{H^\pm G^\pm} 
                \begin{pmatrix} H^{+}\\ G^{+}\end{pmatrix} + \dots.
  \end{split}
\end{align}
with the tadpole coefficients and mass matrices as given in~\cite{Frank:2006yh}.
At lowest order, the tadpoles vanish and the mass matrices
$\mathbf{M}_{ h H A G}^{(0)} = \mathrm{diag} {\left( m_h^2, m_H^2, m_A^2, m_G^2 \right) } $,
$\mathbf{M}_{H^\pm G^\pm}^{(0)} = \mathrm{diag} {\left( m_{H^\pm}^2,m_{G^\pm}^2 \right) } $
are diagonal.

\subsection{Mass spectrum beyond lowest order}
At higher order, the entries of the Higgs boson mass matrices are shifted according to the self-energies,
yielding in general mixing of all tree-level mass eigenstates with equal quantum numbers. In
the case of the neutral Higgs bosons the following ``mass matrix'' 
is evaluated at the two-loop level,
%
\begin{align}
  \label{eq:masscorr}
    \mathbf{M}_{ h H A G}^{(2)} (p^2) &= \mathbf{M}_{ h H A G}^{(0)} 
         -
   \mathbf{\hat{\Sigma}}_{h H A G}   ^{(1)} (p^2)
          -
    \mathbf{\hat{\Sigma}}_{h H A G}  ^{(2)} (0) \, .
\end{align}
Therein,  $ \mathbf{\hat{\Sigma}}_{h H A G} ^{(k)}$ 
denotes the matrix of the renormalized diagonal and non-diagonal 
self-energies for the $h, H, A, G$ fields at loop order $k$.
The present approximation for the  two-loop part 
yielding the leading contributions from the Yukawa sector,
treats the two-loop self-energies at \mbox{$p^2=0$} for the external momentum (as done also for the
leading two-loop $\mathcal{O}{\left(\alpha_t \alpha_s\right)}$ contributions~\cite{Heinemeyer:2007aq})
and neglects contributions from the gauge sector (gaugeless limit).
Furthermore, also the Yukawa coupling of the bottom quark is neglected by
setting the $b$-quark mass to zero.
The diagrammatic calculation of the self-energies has been performed with
{\tt FeynArts}~\cite{Hahn:2000kx} for the generation of the Feynman diagrams 
and {\tt TwoCalc}~\cite{Weiglein:1993hd} for 
the two-loop  tensor reduction and trace evaluation.
The renormalization constants 
have been obtained with the help of {\tt FormCalc}~\cite{Hahn:1998yk}. 

In order to obtain the physical Higgs-boson masses from the dressed propagators
in the considered approximation,
it is sufficient to derive explicitly the entries of the
3$\times$3 submatrix of Eq.~\eqref{eq:masscorr} corresponding to the
$(hHA)$ components. Mixing with the unphysical Goldstone boson yields subleading
two-loop contributions; also Goldstone--$Z$ mixing occurs in principle,
which is related to the other Goldstone mixings by 
Slavnov-Taylor identities~\cite{Baro:2008bg,Williams:2011bu} and 
of subleading type as well~\cite{Hollik:2002mv}. 
However, $A$--$G$ mixing has to be taken into account 
in intermediate steps for a consistent renormalization.

The masses of the three neutral Higgs bosons $h_1,h_2,h_3$,
including the new $\mathcal{O}{\left(\alpha_{t}^{2}\right)}$ contributions, are given by the real parts of the
poles of the $hHA$ propagator matrix, obtained as the zeroes of the determinant of the 
renormalized  two-point vertex function, 
\begin{align}
  \label{eq:higgspoles}
   \operatorname{det}\hat{\Gamma}_{hHA}{\left(p^2\right)} &= 0, &
   \hat{\Gamma}_{hHA}{\left(p^2\right)} &= \I \left[p^2 {\unity} - \mathbf{M}_{hHA}^{(2)}{\left(p^2\right)}\right],
\end{align}
involving the corresponding 3$\times$3 submatrix of Eq.~\eqref{eq:masscorr}.

\subsection{Two-loop renormalization\label{sec:higgsren}}
For obtaining the renormalized self-energies~(\ref{eq:masscorr}),
counterterms have to be introduced for the mass matrices and tadpoles in 
Eq.~(\ref{eq:HiggsPotential}) up to second order in the loop expansion,
\begin{subequations}
\label{eq:counterterms}
\begin{align}
   &  \mathbf{M}_{ h H A G}  
    \, \rightarrow \,
    \mathbf{M}_{ h H A G}^{(0)} +\, \delta^{(1)} \mathbf{M}_{ h H A G}  
    +\, \delta^{(2)} \mathbf{M}_{ h H A G}  \, ,  \\
   & \mathbf{M}_{H^\pm G^\pm}  
    \, \rightarrow \,
   \mathbf{M}_{H^\pm G^\pm}^{(0)}  +\, \delta^{(1)} \mathbf{M}_{H^\pm G^\pm}   
    +\, \delta^{(2)} \mathbf{M}_{H^\pm G^\pm}   \, ,    \\
    & T_i  \, \rightarrow \,   T_i + \,\delta^{(1)} T_i  +\,
    \delta^{(2)} T_i \, , \quad  i=h,H,A, G \, ,
\end{align}
\end{subequations}
as well as field renormalization constants
$  Z_{\mathcal{H}_{i}}  = 1 + \delta^{(1)}Z_{\mathcal{H}_{i}} +
\delta^{(2)}Z_{\mathcal{H}_{i}} $,
which are introduced up to two-loop order for each of the scalar doublets in Eq.~\eqref{eq:Higgsfields} 
through
\begin{align}
  \mathcal{H}_{i} &\rightarrow \sqrt{Z_{\mathcal{H}_{i}}}\mathcal{H}_{i} = \left[ 1 + \frac{1}{2}\delta^{(1)}Z_{\mathcal{H}_{i}} + \frac{1}{2}\delta^{(2)}Z_{\mathcal{H}_{i}} - \frac{1}{8}\left(\delta^{(1)}Z_{\mathcal{H}_{i}}\right)^{2} 
  \right]   \mathcal{H}_{i} \, .
\end{align}
They can be transformed into (dependent) field-renormalization
constants for the mass eigenstates in Eq.~\eqref{eq:higgsmixing}
 according to
\begin{subequations}
\label{eq:neutralhiggsfieldren}
\begin{alignat}{4}
  \begin{pmatrix}h \\ H \end{pmatrix} &\rightarrow
  D(\alpha)\begin{pmatrix}\sqrt{Z_{\mathcal{H}_{1}}} & 0\\ 0 & \sqrt{Z_{\mathcal{H}_{2}}}\end{pmatrix} D(\alpha)^{-1} \begin{pmatrix} h \\ H \end{pmatrix}
   &&\equiv \mathbf{Z}_{hH} \begin{pmatrix}h\\ H \end{pmatrix},\\[0.2cm]
  \begin{pmatrix}A\\ G \end{pmatrix} &\rightarrow D(\beta)\begin{pmatrix}\sqrt{Z_{\mathcal{H}_{1}}} & 0\\ 0 & \sqrt{Z_{\mathcal{H}_{2}}}\end{pmatrix} D(\beta)^{-1} \begin{pmatrix}A\\ G \end{pmatrix}
   &&\equiv \mathbf{Z}_{AG}  \begin{pmatrix}A\\ G\end{pmatrix},\\[0.2cm]
  \begin{pmatrix}H^{\pm}\\ G^{\pm}\end{pmatrix} &\rightarrow D(\beta)\begin{pmatrix}\sqrt{Z_{\mathcal{H}_{1}}} & 0\\ 0 & \sqrt{Z_{\mathcal{H}_{2}}}\end{pmatrix} D(\beta)^{-1} \begin{pmatrix}H^{\pm}\\ G^{\pm}\end{pmatrix}
   &&\equiv  \mathbf{Z}_{H^{\pm}G^{\pm}}  \begin{pmatrix}H^{\pm}\\  G^{\pm} \end{pmatrix} .
\end{alignat}
\end{subequations}

\begin{figure}[t]
  \centering
  \epsfxsize=\textwidth\epsfbox{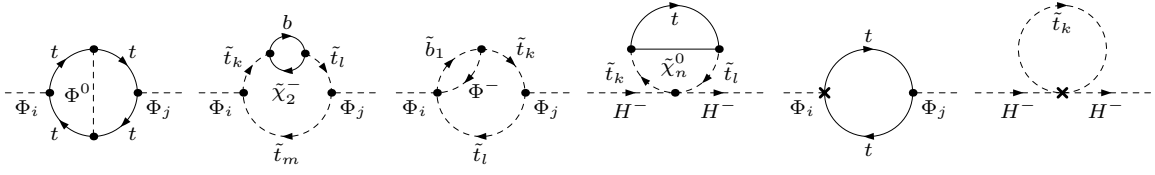}
  \caption{\label{fig:selfenergies}Examples of two-loop self-energy diagrams.
   The cross denotes a one-loop counterterm insertion.
 \mbox{$\Phi_{i} = h,H,A$}; \mbox{$\;\Phi^0 = h,H,A,G$};  \mbox{$\;
   \Phi^- =H^-,G^-$}.  }
\end{figure}

\medskip
At the one-loop level, the  
expressions for the counterterms and for the renormalized self-energies
 $\mathbf{\hat{\Sigma}}_{h H A G} ^{(1)} (p^2)$
are listed in~\cite{Frank:2006yh}. 
For the leading two-loop contributions, we have to evaluate the renormalized 
two-loop self energies at zero external momentum.
In compact matrix notation they can be written as follows,
\begin{align}
\label{eq:renselfenergies}
\mathbf{\hat{\Sigma}}_{h H A G} ^{(2)} (0)  
   & =
\mathbf{\Sigma}_{h H A G} ^{(2)} (0)  -\delta^{(2)} \MZ_{hHAG} \, .
\end{align}
Thereby,  $\mathbf{\Sigma}_{h H A G} ^{(2)}$ denotes
the unrenormalized self-energies corresponding to
the genuine $2$-loop diagrams and diagrams with sub-renormalization,
as illustrated in Fig.~\ref{fig:selfenergies}. 
The quantities in $\delta^{(2)} \MZ_{hHAG}$ can be obtained
as the two-loop content of  the  expression 
\begin{align}
\label{eq:fieldcorr}
  \delta^{(2)}\MZ_{hHAG}
  &= 
 \begin{pmatrix} \mathbf{Z}_{hH}^T & \zero \\
    \zero  & \mathbf{Z}_{AG}^T \end{pmatrix} 
 \left( \mathbf{M}_{hHAG}^{(0)} + \delta^{(1)}\mathbf{M}_{hHAG}  + \delta^{(2)}\mathbf{M}_{hHAG} \right)
  \begin{pmatrix} \mathbf{Z}_{hH} & \zero \\ 
  \zero  & \mathbf{Z}_{AG} \end{pmatrix} .
\end{align}
Explicit formulae will be given in a forthcoming paper.

\begin{figure}[t]
  \centering
  \epsfxsize=\textwidth\epsfbox{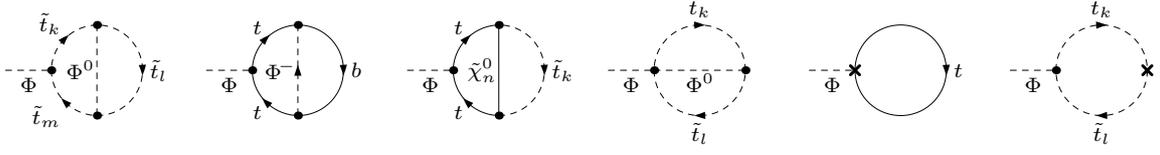}
  \caption{\label{fig:tadpoles}Examples of two-loop tadpole diagrams 
  contributing to $T^{(2)}_i$.
  The cross denotes a one-loop counterterm insertion.
 \mbox{$\Phi_{i} = h,H,A$}; \mbox{$\;\Phi^0 = h,H,A,G$};  \mbox{$\;\Phi^- =H^-,G^-$}. }
\end{figure}

\medskip
The entries of the counterterm matrices in Eq.~\eqref{eq:renselfenergies} 
are determined via renormalization conditions that are extended from
the one-loop level, as specified in~\cite{Frank:2006yh},  to two-loop order:
\begin{itemize}
  \item 
The tadpole counterterms
    $\delta^{(k)} T_i $ 
are fixed by requiring the minimum of the Higgs potential not shifted, 
{\it i.\,e.}\footnote{The counterterms $\delta^{(k)}T_G$ are not independent and do not need separate renormalization conditions}
\begin{subequations}
   \begin{align}
      T_{i}^{(1)} + \delta^{(1)}T_{i}  &= 0, \qquad
      T_{i}^{(2)} + \delta^{(2)}T_{i} + \delta^{(2)}\TZ_{i}  \, = 0,
      \qquad i=h,H,A,
   \end{align}
 with
   \begin{align}
   \left(\delta^{(2)}\TZ_h,\delta^{(2)}\TZ_H \right)
      &= 
    \left(\delta^{(1)} T_h,  \delta^{(1)} T_H\right) \mathbf{Z}_{hH},\\
    \left(\delta^{(2)}\TZ_A,\delta^{(2)}\TZ_G\right)
      &= 
    \left(\delta^{(1)} T_A,  \delta^{(1)} T_G \right) \mathbf{Z}_{AG},
  \end{align}
\end{subequations}
  where only the one-loop parts of the $\mathbf{Z}_{ij}$ from Eq.~\eqref{eq:neutralhiggsfieldren}
  are involved. 
   $T_{i}^{(k)}$ denote the unrenormalized one-point vertex functions;
   two-loop diagrams contributing to $T_i^{(2)}$ are displayed in Fig.~\ref{fig:tadpoles}.
  \item 
The charged Higgs-boson mass $m_{H^\pm}$ is the only independent mass
parameter of the Higgs sector and is used as an input quantity. Accordingly, 
the corresponding mass counterterm is fixed by an independent renormalization condition,
chosen as on-shell condition, which in the \mbox{$p^2=0$} approximation is given by
$ \Real{\hat{\Sigma}_{H^{\pm}}^{(k)}(0)} = 0 $
for the renormalized charged-Higgs self-energy, at the two-loop level specified
in terms of the unrenormalized charged self-energies and respective counterterms,
\begin{subequations}
\begin{align}
\label{eq:renchargedselfenergy}
& \hat{\Sigma}_ {H^\pm} ^{(2)}  \, = \,
\left( \mathbf{\hat{\Sigma}}_{H^\pm G^\pm}^{(2)}  \right) _{11}, \qquad 
 \mathbf{\hat{\Sigma}}_{H^\pm G^\pm}^{(2)} (0)
  \, = \,
\mathbf{\Sigma}_{H^\pm G^\pm} ^{(2)} (0) -\delta^{(2)} \MZ_{H^\pm G^\pm} , \\
& \delta^{(2)}\MZ_{H^{\pm}G^{\pm}}  \,  = \,
      \mathbf{Z}_{H^{\pm}G^{\pm}}^{T} 
      \left( \mathbf{M}_{H^{\pm}G^{\pm}}^{(0)}
        + \delta^{(1)}\mathbf{M}_{H^{\pm}G^{\pm}} 
        + \delta^{(2)}\mathbf{M}_{H^{\pm}G^{\pm}}  \right) 
      \mathbf{Z}_{H^{\pm}G^{\pm}}  , 
\end{align}
\end{subequations}
where only the two-loop content of the last expression is taken.
From the on-shell condition, the independent mass counterterm 
$\delta^{(2)} m_{H^{\pm}}^{2} = 
\left( \delta^{(2)} \mathbf{M}_{H^\pm G^\pm} \right)_{11}$
can be extracted.
  \item 
The field-renormalization constants of the  Higgs mass eigenstates in Eq.~\eqref{eq:neutralhiggsfieldren} 
are combinations of the basic doublet-field renormalization constants
$\delta^{(k)}Z_{\mathcal{H}_{1}}$ and $\delta^{(k)}Z_{\mathcal{H}_{2}}$ $(k=1,2)$, 
which are fixed by the $\overline{DR}$ conditions for the derivatives
of the corresponding self-energies,
\begin{align}
\delta^{(k)}Z_{\mathcal{H}_{1}} & = 
 - \left[ \Sigma^{(k)\, '}_{HH}(0) \right]_{\alpha=0}^{\rm div} , \qquad
\delta^{(k)}Z_{\mathcal{H}_{2}} \, =
 - \left[ \Sigma^{(k)\, '}_{hh}(0) \right]_{\alpha=0}^{\rm div} .
\end{align}
  \item 
\mbox{$t_\beta \equiv \tan\beta$} is
renormalized in the $\overline{DR}$ scheme, 
which has been shown to be a very convenient choice~\cite{Freitas:2002um} 
(alternative process-dependent definitions and renormalization
of  $t_{\beta}$ can be found in Ref.~\cite{Baro:2008bg}). 
It has been clarified in Ref.~\cite{Sperling:2013eva,Sperling:2013xqa} 
that the following identity applies for the $\overline{DR}$ counterterm 
at one-loop order 
and within our  approximations also at the two-loop level:
    \begin{align}
      \delta^{(k)}t_{\beta} &= \frac{1}{2} t_{\beta} \left(\delta^{(k)}Z_{\mathcal{H}_{2}} - \delta^{(k)}Z_{\mathcal{H}_{1}}\right).
    \end{align}

  \item 
In the on-shell scheme, also the counterterms 
$\left.\delta M_W^2\middle/M_W^2\right.$ and $\left.\delta M_Z^2\middle/M_Z^2\right.$ 
are required for renormalization of the
top Yukawa coupling~$h_{t} = \left.\left(e m_{t}\right)\middle/\left(\sqrt{2}s_{\beta}s_{\rm w}M_{W}\right)\right.$.
In the gaugeless limit these ratios have remaining finite and divergent contributions 
arising from the Yukawa couplings, which have to be included as one-loop quantities
\mbox{$\sim h_t^2$}; they are evaluated from the~$W$~and~$Z$ self-energies yielding 
    \begin{align}
       \frac{\delta M_W^2}{M_W^2} &=  \frac{\Sigma_W(0)}{M_W^2} , &
       \frac{\delta M_Z^2}{M_Z^2} &=  \frac{\Sigma_Z(0)}{M_Z^2} , &
       \delta s_{\rm w}^2 &= c_{\rm w}^{2}  \left(\frac{\delta M_{Z}^2}{M_{Z}^2} - \frac{\delta M_{W}^2}{M_{W}^2}\right).
    \end{align}
In the Yukawa approximation, $\delta s_{\rm w}^2 $ is finite. 
The corresponding Feynman graphs are depicted in Fig.~\ref{fig:RCCha}.
\end{itemize}

As a consequence of applying $\overline{DR}$ renormalization conditions the result depends explicitly on the renormalization scale. Conventionally it is chosen as~$m_{t}$ being the default value in {\tt FeynHiggs}.

\medskip
The appearance of~$\delta s_{\rm w}^2$ in the~$\mathcal{O}{\left(\alpha_{t}^{2}\right)}$ terms, as specified above, is a consequence of the on-shell scheme
where the top-Yukawa coupling~$h_{t} = m_{t} / v_{2} = m_{t} /(v s_{\beta})$ is expressed in terms of 
\begin{align}
  \frac{1}{v}  &= \frac{g_2}{\sqrt{2} M_W} = \frac{e}{\sqrt{2} s_{\rm w} M_{W}} .
\end{align}
Accordingly, the one-loop self-energies have to be parametrized in terms
of this representation for~$h_t$ when added to the two-loop
self-energies in Eq.~\eqref{eq:masscorr}. On the other hand, if the Fermi constant~$G_{\rm F}$
is used for parametrization of the one-loop self-energies, the relation
\begin{align}
  \sqrt{2}G_{\text{F}} &= \frac{e^2}{4 s^2_{\rm w} M^2_{W}}\left(1 + \Delta^{(k)}r\right) ,
\end{align}
has to be applied, which gets loop contributions also in the gaugeless limit,
at one-loop order given by
\begin{align}
  \Delta^{(1)} r = - \frac{c^2_{\rm w}}{s^2_{\rm w}} 
  \left(\frac{\delta M_{Z}^2}{M_{Z}^2} - \frac{\delta M_{W}^2}{M_{W}^2} \right)
  \, =\, - \frac{\delta s_{\rm w}^2}{s^2_{\rm w}} \, .
\end{align}
This finite shift in the one-loop self-energies induces two-loop 
$\mathcal{O}{\left(\alpha_t^2\right)}$ terms and has to be taken into account,
effectively cancelling all occurrences of $\delta s_{\rm w}^2$.

\section{Colored-sector input and renormalization}
The two-loop top Yukawa coupling contributions to the self-energies and tadpoles 
involve insertions of counterterms that arise from one-loop renormalization 
of the top and scalar top~$\big(\st\big)$ as well as scalar bottom~$\big(\sbottom\big)$ sectors.
The stop and sbottom mass matrices in the 
$\big(\tilde{t}_{\text{L}},\tilde{t}_{\text{R}}\big)$ and
$\big(\tilde{b}_{\text{L}},\tilde{b}_{\text{R}}\big)$ bases are given by
\begin{align}
  \label{eq:squarks}
    \mathbf{M}_{\tilde{q}} &= 
    \begin{pmatrix}
     m_{\tilde{q}_{\text{L}}}^{2} + m_{q}^{2} + M_Z^2 c_{2\beta} (T_q^3-Q_q s^2_{\rm w}) & 
     m_{q}\left(A_{q}^{*} - \mu\kappa_q \right)\\[0.1cm]
     m_{q}\left(A_{q} - \mu^{*}\kappa_q \right) & 
     m_{\tilde{q}_{\text{R}}}^{2} + m_{q}^{2} + M_Z^2 c_{2\beta} Q_q s^2_{\rm w}
   \end{pmatrix}, &
     \kappa_t &= \frac{1}{t_{\beta}},& \kappa_b &= t_{\beta} ,
\end{align}
with $Q_q$ and $T^3_q$ denoting charge and isospin of $q=t,b$.
$SU(2)$ invariance requires \mbox{$m_{\tilde{t}_{\text{L}}}^{2} =
  m_{\tilde{b}_{\text{L}}}^{2} \equiv m_{\tilde{q}_{3}}^{2}$}. 
In the gaugeless approximation the $D$-terms vanish in 
both the $\st$ and $\sbottom$ matrices.
Moreover, in our approximation the $b$-quark is treated as massless; 
hence, the off-diagonal entries of the sbottom matrix are zero and the
mass eigenvalues can be read off directly,
\mbox{$m_{\tilde{b}_{1}}^{2} = m_{\sbottom_{\rm L}}^2 = m_{\tilde{q}_{3}}^{2}$},
\mbox{$m_{\tilde{b}_{2}}^{2} = m_{\tilde{b}_{\text{R}}}^{2}$}. 
The stop mass eigenvalues 
can be obtained by performing a unitary transformation,
\begin{align}
  \label{eq:squarkdiag}
  \mathbf{U}_{\tilde{t}}\mathbf{M}_{\tilde{t}}\mathbf{U}_{\tilde{t}}^{\dagger}  &= 
  \mathrm{diag}{\left(m_{\tilde{t}_{1}}^{2}, m_{\tilde{t}_{2}}^{2}\right)}.
\end{align}
Since $A_{t}$ and $\mu$ are complex parameters in general, the unitary
matrix $\mathbf{U}_{\tilde{t}}$  consists of one 
mixing angle $\theta_{\st}$ and one phase $\varphi_{\st}$.

\begin{figure}[hbt]
  \centering
  \epsfxsize=\textwidth\epsfbox{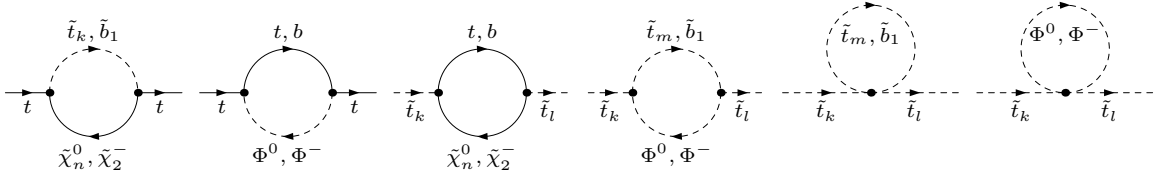}
  \caption{\label{fig:RCTop}
    Feynman diagrams for renormalization of the quark--squark sector.
   \mbox{$\Phi^{0} = h^{0},H^{0},A^{0},G^{0}$}; 
   \mbox{$\Phi^{-} = H^{-},G^{-}$.} }
\end{figure}

\medskip
Five independent parameters are introduced by the quark--squark sector,
which enter the two-loop calculation in addition to those of the previous section: 
the top mass $m_{t}$, the soft SUSY-breaking parameters 
$m_{\tilde{q}_{3}}$ and $m_{\tilde{t}_{\text{R}}}$ ($m_{\tilde{b}_{\text{R}}}$  decouples for $m_{b} = 0$), 
and the complex mixing parameter $A_{t} = \lvert
A_{t}\rvert\E^{i\phi_{A_{t}}}$. 
On top, $\mu$ enters as another free parameter related to the Higgsino sector.  
These parameters have to be renormalized at the one-loop level, 
\begin{align}
  m_t &\rightarrow m_t + \delta m_t, &
  \mathbf{M}_{\st}  &\rightarrow \mathbf{M}_{\st} + \delta \mathbf{M}_{\st} .
\end{align}
The independent renormalization conditions for the colored sector
are formulated in the following way:
\begin{itemize}
  \item 
The mass of the top quark is defined on-shell,  
{\it i.e.}\footnote{$\Realtilde$ denotes the real part of all loop integrals, 
but leaves the couplings unaffected.}
    \begin{align}
      \delta m_{t} &=
      \frac{1}{2} m_{t} \, \Realtilde{\left[\Sigma_{t}^{\text{L}}{\left(m_{t}^{2}\right)} 
    + \Sigma_{t}^{\text{R}}{\left(m_{t}^{2}\right)} + 2\Sigma_{t}^{\text{S}}{\left(m_{t}^{2}\right)}\right]},
    \end{align}
 according to the Lorentz decomposition of the self-energy of the top quark 
 (Fig.~\ref{fig:RCTop})
\begin{align}
\label{eq:Lorentz} 
\Sigma_t (p) & =\,  \not{\! p}\, \omega_-\,  \Sigma_t^{\rm L} (p^2) +
                        \not{\! p}\, \omega_+\,   \Sigma_t^{\rm R}(p^2) 
                        + m_t \,\Sigma_t^{\rm S}(p^2)  
                        + m_t \gamma_5\, \Sigma_t^{\rm PS}(p^2)  .
\end{align}
  \item $m_{\tilde{q}_{3}}^{2}$ and $m_{\tilde{t}_{\text{R}}}^{2}$ are
    traded for $m_{\tilde{t}_{1}}^{2}$ and $m_{\tilde{t}_{2}}^{2}$, 
    which are then fixed by on-shell conditions for the top-squarks,
    \begin{align}
      \delta m_{\tilde{t}_{i}}^{2} &= \Realtilde{\, \Sigma_{\tilde{t}_{ii}}{\left(m_{\tilde{t}_{i}}^{2}\right)}}, 
      \quad  i=1,2 \, ,
   \end{align}
  involving the diagonal $\st_1$ and $\st_2$  self-energies 
  (diagrammatically visualized in Fig.~\ref{fig:RCTop}).
  These on-shell conditions determine the diagonal entries
  of the counterterm matrix
  \begin{align}
   \label{eq:stopcountermmatrix}     
     \mathbf{U}_{\tilde{t}}\delta\mathbf{M}_{\tilde{t}}\mathbf{U}_{\tilde{t}}^{\dagger} & =
     \begin{pmatrix} 
         \delta m_{\tilde{t}_{1}}^{2} & \delta m_{\tilde{t}_{1}\tilde{t}_{2}}^{2}  \\ 
         \delta m_{\tilde{t}_{1}\tilde{t}_{2}}^{2\,*} & \delta m_{\tilde{t}_{2}}^{2} 
    \end{pmatrix} .
    \end{align}
  \item 
   The mixing parameter $A_{t}$ is correlated with the $\st$-mass
    eigenvalues, $t_\beta$, and $\mu$, through Eq.~\eqref{eq:squarkdiag}. 
   Exploiting Eq.~\eqref{eq:stopcountermmatrix}
   and the unitarity of $\mathbf{U}_{\tilde{t}}$ yields the expression
    \begin{multline}
    \label{eq:Atrenormalization}
      \left(A_{t} - \frac{\mu^{*}}{t_{\beta}}\right)\delta m_{t} + m_{t} \left(\delta A_{t} - 
      \frac{\delta\mu^{*}}{t_{\beta}} + \frac{\mu^{*}\delta t_{\beta}}{t_{\beta}^{2}}\right) =\\
      \mathbf{U}_{\tilde{t}\,11}\mathbf{U}_{\tilde{t}\,12}^{*}\left(\delta m_{\tilde{t}_{1}}^{2} - \delta m_{\tilde{t}_{2}}^{2}\right)
      + \mathbf{U}_{\tilde{t}\,21}\mathbf{U}_{\tilde{t}\,12}^{*}\delta m_{\tilde{t}_{1}\tilde{t}_{2}}^{2}
      + \mathbf{U}_{\tilde{t}\,22}\mathbf{U}_{\tilde{t}\,11}^{*}\delta m_{\tilde{t}_{1}\tilde{t}_{2}}^{2\,*}.
    \end{multline}
    For the non-diagonal entry of~(\ref{eq:stopcountermmatrix}),
    the renormalization condition
    \begin{align}
      \delta m_{\tilde{t}_{1}\tilde{t}_{2}}^{2} &= 
      \frac{1}{2}\Realtilde{\left[\Sigma_{\tilde{t}_{12}}{\left(m_{\tilde{t}_{1}}^{2}\right)}
          + \Sigma_{\tilde{t}_{12}}{\left(m_{\tilde{t}_{2}}^{2}\right)}\right]} 
    \end{align}
   is imposed, as in~\cite{Heinemeyer:2007aq}, which involves the 
   non-diagonal $\st_1$--$\st_2$ self-energy (Fig.~\ref{fig:RCTop}). 
   By means of Eq.~\eqref{eq:Atrenormalization} the counterterm $\delta A_t$ 
   is then determined. Actually this yields two conditions, 
   for $\lvert A_t \rvert $ and for the phase $\phi_{A_t}$ separately.
   The additionally required  
   mass counterterm $\delta\mu$ is obtained as described below 
   in section~\ref{sec:higgsinos}.
  \item 
As already mentioned, the relevant sbottom mass is not an independent parameter, 
and hence its counterterm is a derived quantity that can be obtained 
from Eq.~\eqref{eq:stopcountermmatrix},
\begin{multline}
  \delta m_{\tilde{b}_{1}}^{2} \,\equiv\, \delta m_{\tilde{q}_{3}}^{2} =\\  
  \lvert\mathbf{U}_{\tilde{t}\,11}\rvert^{2}\delta m_{\tilde{t}_{1}}^{2} + 
  \lvert\mathbf{U}_{\tilde{t}\,12}\rvert^{2}\delta m_{\tilde{t}_{2}}^{2}
  - \mathbf{U}_{\tilde{t}\,22}\mathbf{U}_{\tilde{t}\,12}^{*}\delta m_{\tilde{t}_{1}\tilde{t}_{2}}^{2}
  - \mathbf{U}_{\tilde{t}\,12}\mathbf{U}_{\tilde{t}\,22}^{*}\delta m_{\tilde{t}_{1}\tilde{t}_{2}}^{2\,*} 
  - 2m_{t}\delta m_{t} .
\end{multline}

\end{itemize}

\section{Chargino--neutralino-sector input and renormalization}
\label{sec:higgsinos}
For the calculation of the $\mathcal{O}{\left(\alpha_{t}^{2}\right)}$ contributions 
to the Higgs boson self-energies and tadpoles,
also the neutralino and chargino sectors have to be considered. 
Chargino/neutralino vertices and propagators enter
only at the two-loop level and thus do not need renormalization;
in the one-loop terms, however, the Higgsino-mass 
parameter $\mu$ appears and the counterterm $\delta\mu$ 
is required for the one-loop subrenormalization.
The mass matrices in the bino/wino/higgsino bases
are given by
\begin{align}
  \mathbf{Y} &= 
  \begin{pmatrix}
   M_{1} & 0 & -M_{Z} s_{\rm w} c_{\beta} & M_{Z} s_{\rm w} s_{\beta}\\ 0 & M_{2} & M_{Z} c_{\rm w} c_{\beta} & M_{Z} c_{\rm w} s_{\beta}\\
   -M_{Z} s_{\rm w} c_{\beta} & M_{Z} c_{\rm w} c_{\beta} & 0 & -\mu\\ M_{Z} s_{\rm w} s_{\beta} & M_{Z} c_{\rm w} s_{\beta} & -\mu & 0 
   \end{pmatrix}
     &\text{and}&&
  \mathbf{X} &= 
  \begin{pmatrix}
    M_{2} & \sqrt{2} M_{W} s_{\beta}\\
    \sqrt{2} M_{W} c_{\beta} & \mu
   \end{pmatrix} .
\end{align}
Diagonal matrices with real and positive entries are obtained with the
help of unitary matrices $ \mathbf{N}, \mathbf{U}, \mathbf{V}$
by the transformations
\begin{align}
 \label{eq:ewdiag}
  \mathbf{N}^{*}\mathbf{Y}\mathbf{N}^{\dagger} &= 
  \mathrm{diag}{\left(m_{\tilde{\chi}^{0}_{1}}, m_{\tilde{\chi}^{0}_{2}}, m_{\tilde{\chi}^{0}_{3}}, m_{\tilde{\chi}^{0}_{4}}\right)}, &
  \mathbf{U}^{*}\mathbf{X}\mathbf{V}^{\dagger} &= \mathrm{diag}{\left(m_{\tilde{\chi}^{\pm}_{1}}, m_{\tilde{\chi}^{\pm}_{2}}\right)}.
\end{align}
In the gaugeless limit the off-diagonal $(2\times2)$ blocks of
$\mathbf{Y}$ and the off-diagonal entries of $\mathbf{X}$ vanish. 
For this special case the transformation matrices and diagonal entries in Eq.~\eqref{eq:ewdiag} simplify,
\begin{subequations}\begin{align}\begin{aligned}
  \mathbf{N} &= 
  \begin{pmatrix}
    \begin{matrix}\E^{\frac{\I}{2} \phi_{M_{1}}} & 0\\ 0 & \E^{\frac{\I}{2} \phi_{M_{2}}}\end{matrix} & 
     \zero \\  \zero  & \frac{1}{\sqrt{2}}\E^{\frac{\I}{2}\phi_{\mu}}
        \begin{pmatrix} 1 & -1\\ \I & \I \end{pmatrix} 
  \end{pmatrix}, &
  \mathbf{U} &= \begin{pmatrix}\E^{\I\phi_{M_{2}}} & 0\\ 0 & \E^{\I\phi_{\mu}}\end{pmatrix}, &
 \mathbf{V} &= \unity \,  ;\end{aligned}\\
 \begin{aligned}
   m_{\tilde{\chi}^{0}_{1}} &= \lvert M_{1}\rvert, & 
   m_{\tilde{\chi}^{0}_{2}} &= \lvert M_{2}\rvert, & 
   m_{\tilde{\chi}^{0}_{3}} &= \lvert\mu\rvert, &
   m_{\tilde{\chi}^{0}_{4}} &= \lvert\mu\rvert, &
   m_{\tilde{\chi}^{\pm}_{1}} &= \lvert M_{2}\rvert, &
   m_{\tilde{\chi}^{\pm}_{2}} &= \lvert\mu\rvert \, ;
\end{aligned}\end{align}\end{subequations}
and only the Higgsinos $\tilde{\chi}^{0}_{3},\,\tilde{\chi}^{0}_{4},\, \tilde{\chi}^{\pm}_{2}$ 
remain in the $\mathcal{O}{\left(\alpha_{t}^{2}\right)}$ contributions.

\begin{figure}[hbt]
  \centering
  \epsfxsize=\textwidth\epsfbox{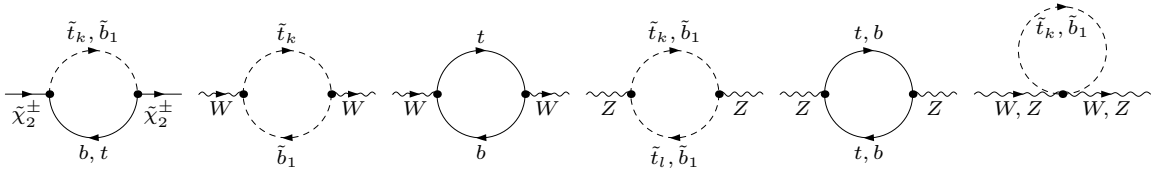}
  \caption{\label{fig:RCCha}
   Feynman diagrams for the counterterms 
  $\delta\mu$, $\delta M_{W}/M_W$, and  $\delta M_{Z}/M_Z$.}
\end{figure}

\medskip
The Higgsino mass parameter $\mu$ is an independent input quantity and
has to be renormalized accordingly, $\mu \rightarrow \mu + \delta\mu$,
fixing the counterterm $\delta\mu$ by an independent renormalization
condition, which renders the one-loop subrenormalization complete. 
Together with the soft-breaking parameters $M_{1}$ and $M_{2}$,  $\mu$ can
be defined  in the neutralino/chargino sector by requiring on-shell
conditions for the two charginos and one neutralino. However, since
only $\delta\mu$ is required here, it is sufficient to impose 
a renormalization condition for $\tilde{\chi}^{\pm}_{2}$ only;
the appropriate on-shell condition reads,
\begin{align}
\label{eq:murenormalization}
  \begin{split}
    \delta\mu &= \frac{1}{2}\E^{\I \phi_{\mu}}  \left\{
    \lvert\mu\rvert \, \Real{ \left[ \Sigma^{\text{L}}_{\tilde{\chi}^\pm_2}(\lvert\mu\rvert^{2})
        + \Sigma^{\text{R}}_{\tilde{\chi}^\pm_2}(\lvert\mu\rvert^{2}) \right]}
      + 2\Real{
        \left[\Sigma^{\text{S}}_{\tilde{\chi}^\pm_2}(\lvert\mu\rvert^{2}) \right]}  \right\} ,
  \end{split}
\end{align}
where the Lorentz decomposition of the self-energy 
for the Higgsino-like chargino $\tilde{\chi}^{\pm}_{2}$ 
(see Fig.~\ref{fig:RCCha})
has been applied, in analogy to Eq.~\eqref{eq:Lorentz}.

Another option implemented in the~$\mathcal{O}{\left(\alpha_{t}^{2}\right)}$ result
is the $\overline{DR}$ renormalization of $\mu$,
which defines the counterterm $\delta\mu$ in the $\overline{DR}$ scheme,
{\it i.e.}~by the divergent part of the expression in Eq.~\eqref{eq:murenormalization}.
For the numerical analysis and comparison with~\cite{Brignole:2001jy} the $\overline{DR}$~scheme is chosen at the scale~$m_{t}$.

\section{Numerical analysis}
In this section we focus on the lightest Higgs-boson mass derived from Eq.~\eqref{eq:higgspoles} 
and present results for $m_{h_1}$ in different parameter scenarios.
In each case the complete one-loop results and the  $\mathcal{O}{\left(\alpha_{t}\alpha_{s}\right)}$ terms 
are obtained from  {\tt FeynHiggs}, while  the $\mathcal{O}{\left(\alpha_{t}^{2}\right)}$ terms
are computed by means of the corresponding two-loop self-energies specified in the previous sections
with the parameters~$\mu, t_{\beta}$ and the Higgs field-renormalization constants defined in the~$\overline{DR}$ scheme at the scale~$m_{t}$.
Thereby, the new $\mathcal{O}{\left(\alpha_{t}^{2}\right)}$ self-energies are
combined with the results of the other available self-energies according to Eq.~\eqref{eq:masscorr} 
within {\tt FeynHiggs}, and the masses are derived via Eq.~\eqref{eq:higgspoles}.
For comparison with previous results~$G_{\text{F}}$ is chosen for normalization as mentioned at the end of section~\ref{sec:higgsren}.

\smallskip
The input parameters for the numerical results in this section are shown in the
figures or the captions, respectively, when they are varied. The residual parameters,
being the same for all the plots, are chosen as \mbox{$M_2 = 200 \text{ GeV}$}, 
\mbox{$M_1 = \left.\left(5s_{\rm w}^2\right)\middle/\left(3c_{\rm w}^2\right) M_2\right.$}, and
\mbox{$m_{\tilde{l}_{\rm L}} =  m_{\tilde{q}_{\rm L}} = m_{\tilde{l}_{\rm R}} = m_{\tilde{q}_{\rm R}} = 2000 \text{ GeV}$}
for the first two sfermion generations, together with the Standard Model input \mbox{$m_t = 173.2 \text{ GeV}$} and \mbox{$\alpha_s = 0.118$}.

\begin{figure}[htb]
  \centering
  \epsfxsize=.9\textwidth\epsfbox{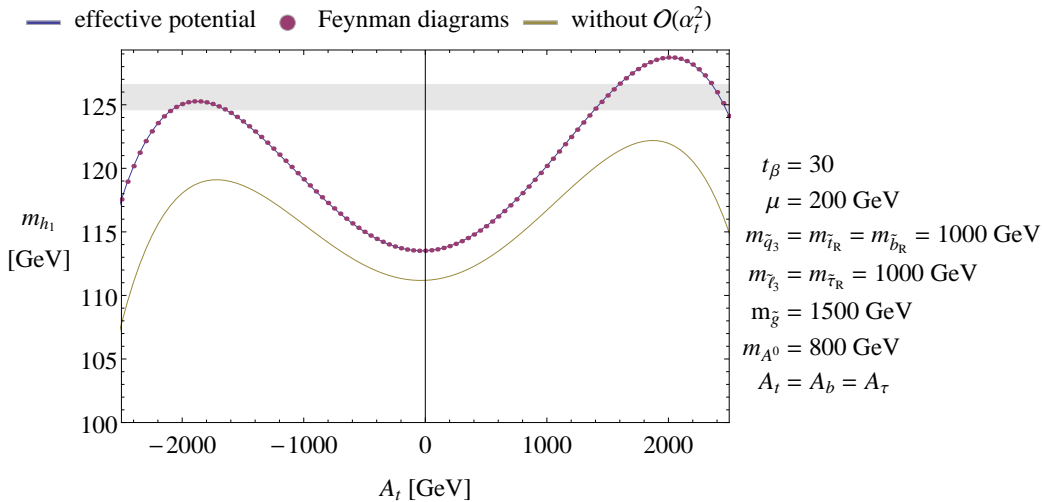}
  \caption{\label{fig:realAtcompare}Comparison of the result for the
    lightest Higgs-boson mass in the effective potential approach
    (blue) and the Feynman-diagrammatic approach (red). The curves are
    lying on top of each other, indicating the agreement of both
    calculations in the limit of real parameters. For reference the
    result without the contributions of
    $\mathcal{O}{\left(\alpha_{t}^{2}\right)}$ is shown (yellow). 
   The grey area depicts the mass range between~$124.6$~GeV and~$126.6$~GeV.}
\end{figure}

\smallskip
As a first application, we study the case of the real MSSM, where an analytic result of 
the $\mathcal{O}{\left(\alpha_{t}^{2}\right)}$ contributions is known~\cite{Brignole:2001jy}
from a calculation making use of the effective-potential method.
The version of {\tt FeynHiggs} for real parameters has this result included, 
making thus a direct comparison with the prediction of our new diagrammatic calculation possible.
Thereby all parameters and the renormalization have been adapted to agree with Ref.~\cite{Brignole:2001jy}.
Very good agreement is found between the two results that
have been obtained in completely independent ways. As an example,
this feature is displayed in Fig.~\ref{fig:realAtcompare}, where the
shift from the~$\mathcal{O}{\left(\alpha_{t}^{2}\right)}$ terms in the
two approaches are shown on top of the mass prediction without those terms. 
The grey band depicts the mass range~$125.6\pm 1$~GeV 
around the Higgs signal measured by ATLAS and CMS.
The mass shifts displayed in Fig.~\ref{fig:realAtcompare}
underline the importance of the two-loop Yukawa contributions for a 
reliable prediction of the  lightest Higgs boson mass.

\begin{figure}
  \centering
  \epsfxsize=.9\textwidth\epsfbox{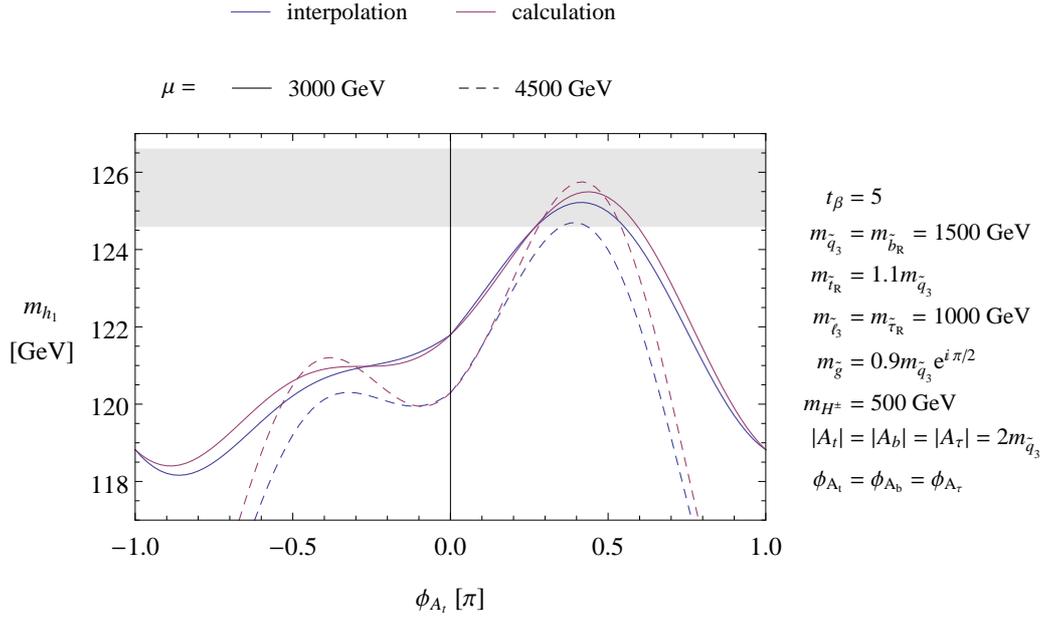}
  \caption{\label{fig:complexPhiAt}
Result from the diagrammatic calculation for complex parameters (red), 
in comparison with the approximate result from interpolation between 
the phases \mbox{$\phi_{A_{t}} = 0,\, \pm\pi$}. 
The grey area depicts the mass range between 124.6 GeV and 126.6 GeV.}
\end{figure}

\smallskip
In the present version of {\tt FeynHiggs} for complex parameters, the dependence of the 
$\mathcal{O}{\left(\alpha_{t}^{2}\right)}$ terms 
on the phases of $\phi_{A_{t}}$ and $\phi_{\mu}$ 
is approximated by an interpolation between the real results 
for the phases~$0$ and~$\pm\pi$~\cite{Hahn:2009zz,Hahn:2007fq}.
A comparison with the full diagrammatic calculation 
yields deviations that can be notable, in particular for large~$|A_t|$. 
Fig.~\ref{fig:complexPhiAt} displays the quality of the interpolation as a function
of~$\phi_{A_t}$ and shows that the deviations become more pronounced with rising~$\mu$, which is kept real.
[Also the admixture of the $CP$ odd part in $h_1$ is increasing with $\mu$, 
but it is in general small, below $2\%$.] 
The asymmetric behaviour with respect to $\phi_{A_{t}}$ is caused
by the phase of the gluino mass in the
$\mathcal{O}{\left(\alpha_{t}\alpha_{s}\right)}$ contributions. 
The shaded area again illustrates the interval~$\left[124.6, 126.6\right]$~GeV.

\begin{figure}[t]
  \centering
  \epsfxsize=\textwidth\epsfbox{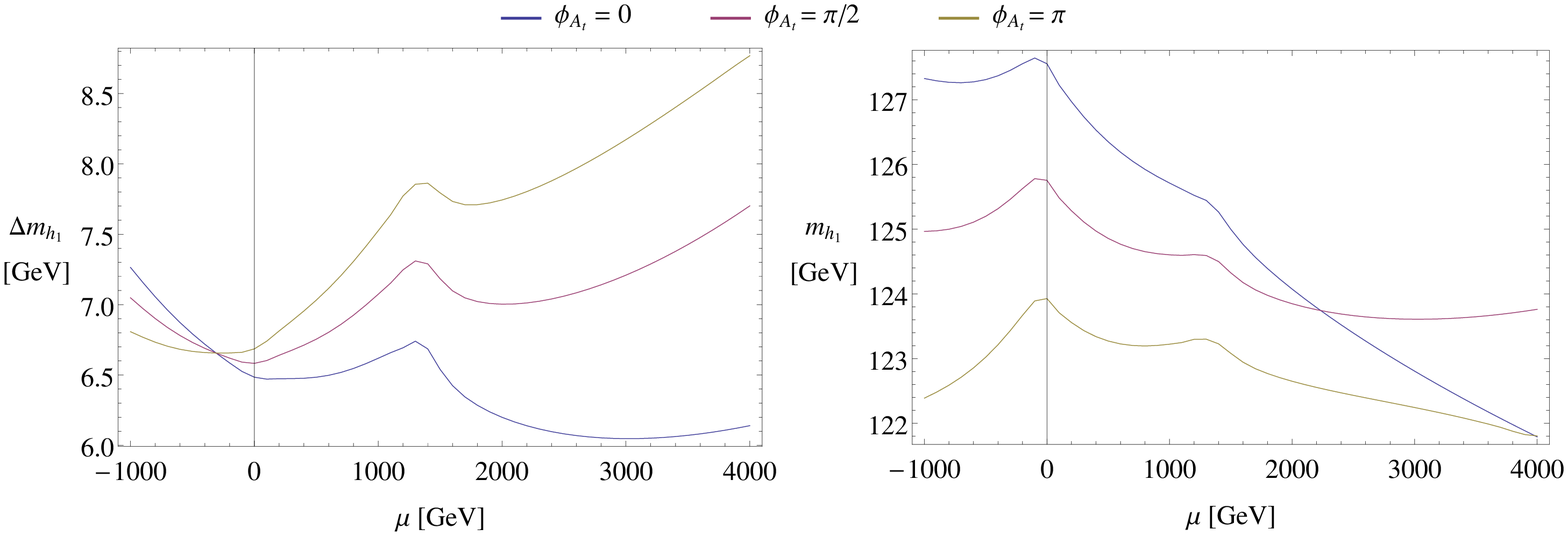}
  \caption{\label{fig:complexMu}
Left: Increasing phase dependence of the
$\mathcal{O}{\left(\alpha_{t}^{2}\right)}$ contributions 
to the lightest Higgs-boson mass with rising $\mu$. 
Right: The lightest Higgs-boson mass including all known contributions.  
The parameters are chosen as follows: 
\mbox{$t_{\beta} = 7$}, 
\mbox{$m_{\tilde{q}_{3}} = m_{\tilde{t}_{\text{R}}} = m_{\tilde{b}_{\rm R}} = 1500 \text{ GeV}$}, 
\mbox{$m_{\tilde{g}} = 1500 \text{ GeV}$}, 
\mbox{$m_{H^{\pm}} = 500 \text{ GeV}$}, 
\mbox{$A_{t} = A_{b} = A_\tau = 1.6 m_{\tilde{q}_{3}}$},
\mbox{$m_{\tilde{l}_3} = m_{\tilde{\tau}_{\rm R}} = 1000 \text{ GeV}$}.
}
\end{figure}

\smallskip
In Fig.~\ref{fig:complexMu} the dependence on $\mu$ is shown for the mass shift 
originating from the $\mathcal{O}{\left(\alpha_{t}^{2}\right)}$ terms 
and for the full result for the lightest Higgs-boson mass, choosing
different values for the phase~$\phi_{A_{t}}$. 
Particularly for large $\mu$ the results can vary significantly and lead to different predictions 
for the lightest Higgs mass. 
The kinks around $\mu \approx 1200$ GeV and $\mu \approx 1450$ GeV 
arise from physical thresholds of the decay of a stop into a higgsino and top.

\section{Conclusions}
We have presented new results for the two-loop Yukawa contributions
$\mathcal{O}{\left(\alpha_{t}^{2}\right)}$ from the top--stop sector
in the calculation of the Higgs-boson masses
of the MSSM with complex parameters. They generalize the previously
known result for the real MSSM to the case of complex phases entering
at the two-loop level; in the limit of real parameters they confirm the
previous result. Combining the new terms with the existing one-loop 
result and leading two-loop terms of $\mathcal{O}{\left(\alpha_t \alpha_s\right)}$    
yields an improved prediction for the Higgs-boson mass spectrum also 
for complex parameters that is equivalent in accuracy to that of
the real MSSM. In the numerical discussion we have focused on the mass
of the lightest neutral boson, $m_{h_1}$, which receives special interest 
by comparison with the mass of the recently  discovered Higgs signal.
The mass shifts originating from the $\mathcal{O}{\left(\alpha_{t}^{2}\right)}$ 
terms are significant, and hence an adequate treatment also for complex parameters
is an obvious requirement.
The new terms will be included in the code {\tt FeynHiggs}, where so far
the complex phases are treated in an approximate way by interpolating
between the real results for phases $0$ and $\pm \pi$.
A more elaborate discussion of the Higgs-boson masses and mixings 
including the heavier states $h_2, \, h_3$ will be given in a forthcoming publication.

\clearpage

\section*{Acknowledgement}
\sloppy{We thank Stefano Di Vita, Thomas Hahn, Sven Heinemeyer, Heidi Rzehak, Pietro Slavich,
\mbox{Alexander Voigt}, and Georg Weiglein for helpful discussions
and valuable support.}

\end{document}